\documentclass[preprint,5p,times]{elsarticle}

\usepackage[utf8]{inputenc} 
\usepackage[T2A]{fontenc}
\usepackage[english]{babel}   
\usepackage{lineno}
\usepackage{graphicx}
\usepackage{xcolor}
\usepackage[pdftex, backref, colorlinks]{hyperref}
\usepackage{multirow}
\usepackage{enumitem}
\usepackage{amsmath}
\modulolinenumbers[5]

\journal{Moscow University Physics Bulletin}

\usepackage{fancyhdr}
\begin{document}

%\linenumbers
\pagestyle{fancy}

\begin{frontmatter}
\title{Application of convolutional neural networks \\for extensive air shower separation \\in the SPHERE-3 experiment\tnoteref{t1}} 

\tnotetext[t1]{Published as a conference paper at DLCP2024, June 19-21, 2024, Moscow, Russia. \url{https://dlcp2024.sinp.msu.ru}}

\address[addressa]{Skobeltsyn Institute of Nuclear Physics Lomonosov Moscow State University}
\address[addressb]{Lomonosov Moscow State University, Faculty of Physics}
\address[addressc]{Lomonosov Moscow State University, Faculty of Computational Mathematics and Cybernetics}
\address[addressd]{Lomonosov Moscow State University, Department of Cosmic Space}

\author[addressa]{E.~L.~Entina\corref{correspondingauthor1}}
\cortext[correspondingauthor1]{Corresponding author}
\ead{el.entina@physics.msu.ru}
\author[addressa,addressb]{D.~A.~Podgrudkov}
\ead{d.a.podgrudkov@physics.msu.ru}

\author[addressa,addressb]{C.~G.~Azra}
\author[addressa]{E.~A.~Bonvech}
\author[addressa,addressd]{O.~V.~Cherkesova}
\author[addressa]{D.~V.~Chernov}
\author[addressa,addressb]{V.~I.~Galkin}
\author[addressa,addressb]{V.~A.~Ivanov}
\author[addressa,addressb]{T.~A.~Kolodkin}
\author[addressa,addressb]{N.~O.~Ovcharenko}
\author[addressa]{T.~M.~Roganova}
\author[addressa,addressc]{M.~D.~Ziva}

\begin{abstract}
A new SPHERE-3 telescope is being developed for cosmic rays spectrum and mass composition studies in the 5--1000~PeV energy range. Registration of extensive air showers using reflected Cherenkov light method applied in the SPHERE detector series requires a good trigger system for accurate separation of events from the background produced by starlight and airglow photons reflected from the snow. Here we present the results of convolutional networks application for the classification of images obtained from Monte Carlo simulation of the detector. Detector response simulations include photons tracing through the optical system, silicon photomultiplier operation and electronics response and digitization process. The results are compared to the SPHERE-2 trigger system performance.

\end{abstract}

%    07.05.Mh, 07.05.Pj, 96.50.S--, 96.50.sd}\par 

%DOI:  
\end{frontmatter}

\section{Introduction}
\label{intro}
\begin{figure}[tbp]
    \includegraphics{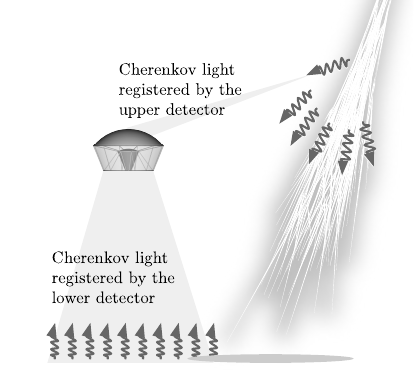}% 
    \caption{ SPHERE-3 experiment scheme.}
    \label{fig:img_01}
\end{figure}

Cherenkov light (CL) is the optical component of extensive air showers (EAS) that allows to obtain some of information about the primary particle of the shower and therefore about cosmic rays.
For EAS detectors, in order to increase the number of registered EAS events, there is a need to register CL from as large an area as possible. In the case of ground experiments~\cite{TAIGA,Yakutsk,MSU} this involves an increase in the number of detectors. An alternative is the method of EAS registration, which is used in the SPHERE experiment series. 
The telescope is risen to a certain altitude above the snow surface and detects CL reflected from the snow (Fig.~\ref{fig:img_01}). This allows to use only one compact detector as well as to change the resolution in the energy spectrum sections by changing the telescope altitude.
%The SPHERE experiments aimed at cosmic ray studies near the ``knee'' region (1--100~PeV) are based on the reflected Cherenkov light (CL) from extensive air showers (EAS) registration method. %The idea of this technique was first proposed by A.E. Chudakov~\cite{Chu1972}.
Previous experiments (SPHERE-1 and SPHERE-2) are described in detail in~\cite{Antonov2015b}. 

Comparing to the SPHERE-2 detector the new one will have a larger aperture and higher spacial resolution (2653 pixels in the light sensitive camera instead of 109), and some other features (like direct EAS CL registration, see~\cite{Bonvech2024}). Since the new detector will be carried by an unmanned aerial vehicle its weight should be smaller than that of SPHERE-2. The best way to cut down the detector weight is to switch from photomultiplier tubes (PMT) to silicon photomultipliers (SiPM). SiPMs find more and more use in astrophysics applications, specifically in imaging air Cherenkov telescopes cameras~\cite{FACT,MAGIC,CTA}, since they are smaller, lighter, and operate at lower voltages than PMTs. Lower operation voltages inherently allow to save more weight on power supply units, what in turn, reduces power consumption and saves on batteries' weight.

But SiPMs have some specific properties that affect measurements: high temperature dependence of both amplification and sensitivity, and optical cross-talk. The former can be resolved by camera and power supply electronics temperature control and stabilization, which is achievable, since power consumption and resulting heat generation are low. The latter, however, should be taken into account in detector design and data analysis. A SiPM is a set of a large number of avalanche photodiodes (individually referred to as microcells) working in Geiger mode that are combined on a single crystal. When a photon hits a microcell it discharges  and this discharge itself can produce optical photons (as does any current in a p-n junction), which can trigger other microcells in the SiPM. This process is called "optical cross-talk" (or simply "cross-talk" in short).

Since the cross-talk is, essentially, a random process and the number of microcells in an individual SiPM is large (from a few hundred to tens of thousands) the actual number of microcells triggered by a single photon and, thus, the resulting response amplitude, is also random. But while the probability of a cross-talk may be quite high (up to 0.35 and above depending on the SiPM overvoltage), its effect is always an additional signal above the expected and can be accounted for during the calibration and data analysis procedure. But for the trigger system the situation is different.

\section{Problem formulation}
\label{sec:problem}
Since the telescope receives light continuously, it is necessary to separate the signal from the noise in order to start event recording. In the previous version of the telescope, this was done by a topological trigger (described below). Because of the SiPM cross-talks, this approach will not work as intended in the new design. We propose a two-stage trigger (described in~\ref{sec:method}, the mathematical formulation is also given) that allows to circumvent this problem.

\begin{figure}[tbp]
\includegraphics[width=.47\textwidth]{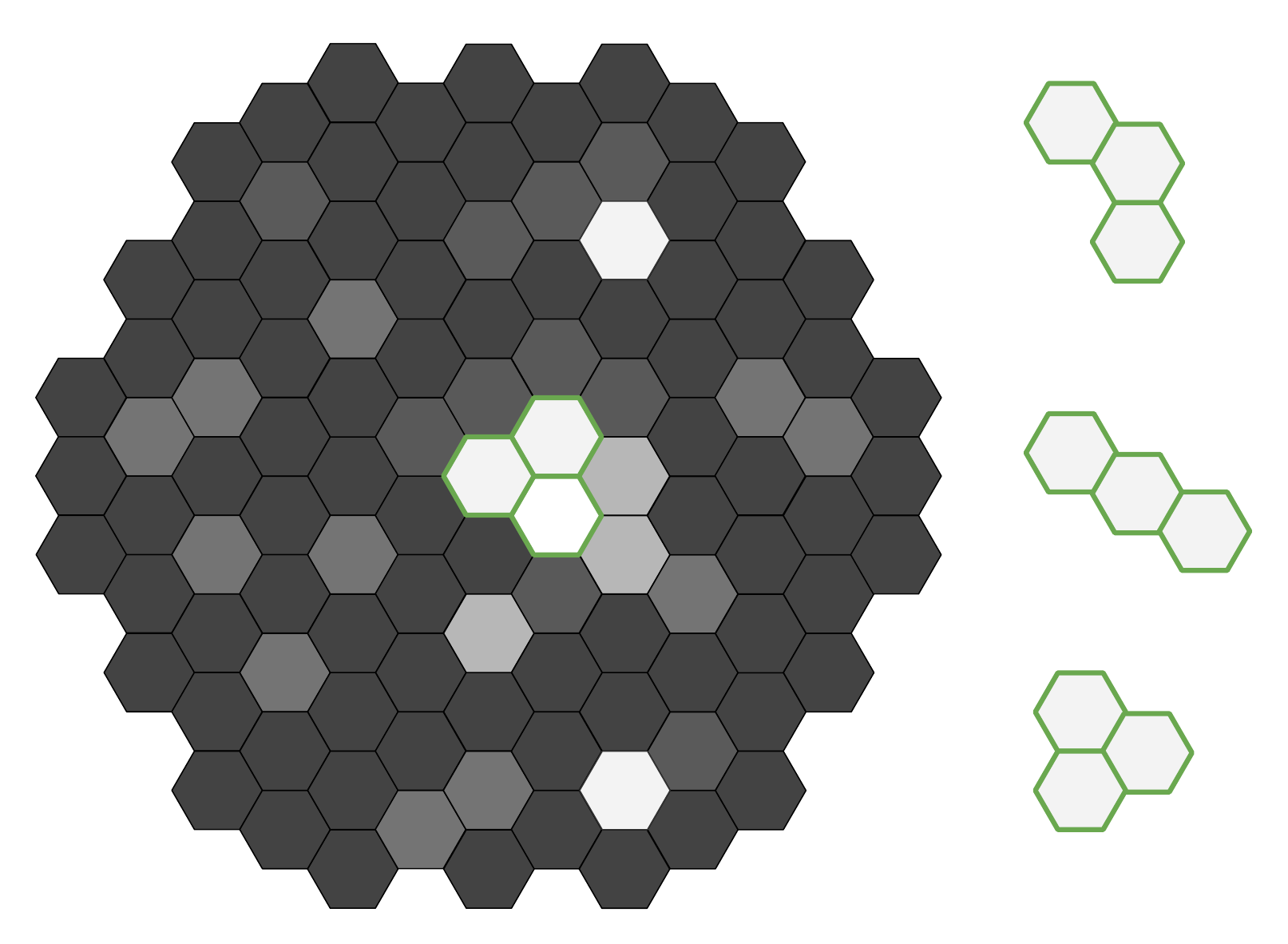}% 
\caption{\label{fig:img_02} Example of a topologiсal trigger activation. On the right are shown all possible triplet variations (without rotations).}
\end{figure}

The SPHERE-2 detector, when triggered, began recording the incoming photon flow. The trigger was topological: first, adaptive thresholds were calculated depending on the noise level (which made the trigger independent of background illumination), after which the topological logic was triggered: if 3 adjacent pixels (Fig.~\ref{fig:img_02}) on the mosaic exceeded the threshold within certain time window, recording began.

It is impossible to apply this approach to the new type of pixels. SiPMs are much more susceptible to multiple triggering than PMTs. With an increase in the number of pixels, increases the number of false triggerings due to the larger number of adjacent pixels triplets (which is nearly directly proportional to the number of pixels). 

Another issue is that there is a constant flux of background photons and dark current electrons.  PMTs by their design cannot produce more than one photoelectron per photon, but have a wide amplification distribution (see, for example, manufacturer or our data~\cite{Antonov2016,Hamamatsu}). As a result some photons (a few percent) produce a signal amplitude more than twice the average. But chances for high multiplicity amplitudes are negligible, since the main source of amplification variation is the number of secondary electrons emitted from first dynode which follows the Poisson distribution. SiPMs have a narrow individual microcell amplification distribution but their cross-talk can cause the signal to be much higher. Studies of SiPMs proposed for use in SPHERE-3 show that chances for high cross-talk multiplicity $n$ is roughly proportional to $p^n$, where $p$ is the cross-talk probability~\cite{Amineva2023,Rehbein2023}. This means that for a steady stream of background photons the probability that from a single photoelectron a PMT will produce a pulse 10 times higher than average is below $10^{-30}$, while for SiPM with 40\% cross-talks this probability is barely around $10^{-4}$.

The background photon flux $F_b$ for the SPHERE-3 experiment scheme was estimated to be around 0.04~ph./ns per pixel. With SiPM photon detection efficiency taken into account this gives $F=0.013$~ph.e./ns. If the SPHERE-2 procedure will be followed, the thresholds setup procedure will set them unacceptably high. The procedure sets the thresholds individually in each pixel in a few passes, every time checking the pixel activation rate and matching it against the target rate. The first is a top to bottom pass, that lowers the initial \textit{a priori} high threshold by measuring every second the average pixel activation rate. The second pass is individual threshold ``tempering'', when each pixel activation frequency should not exceed the target one (an average over 4 seconds was used), and if it does, the thresholds are increased. The last pass is a general trigger system check, that it does not trigger more often than at a certain rate (1~Hz due to data acquisition system limitations). 

The target activation rate for pixels in the SPHERE-2 was $f_t=100$~Hz. With $\tau=1~\mu$s coincidence scheme gates the chance of random activation for a pair of pixels is $f_t\tau$ and for a triplet $\left(f_t\tau\right)^2$. This gives a random trigger rate (for $N_p=109$ pixels) $N_pf_t^3\tau^2\sim10^{-4}$~Hz --- around once per 3 hours, what is well above required. The SPHERE-2 electronics had low amplification and coarse threshold steps, therefore there was no possible way to further lower the thresholds (e.g. pixels activation rates were either below 1--2~Hz or above 10~kHz).

Application of this logic to SHPERE-3 SiPMs will result in an expected rate of cross-talk multiplicity $n$ per pixel:
\begin{equation}
    f(n)=p^nF.
\end{equation}

This is a simplified approach since cases of simultaneous arrival of 2 or more background photons are ignored. Also cross-talks are treated as simple microcell triggering chains, e.g. a triggered microcell has a chance $p$ to trigger another one, while a correct simulation would be when a triggered microcell has a chance to trigger a number of other microcells with a Poisson mean $\lambda=\ln(1-p)^{-1}$ (that should also depend on microcell position in the SiPM). A simplified model yields lower frequency but the difference is not very significant at higher $n$ (a more accurate probability function can be found in~\cite{Rehbein2023}). 

For the target rate $f_t=100$~Hz $n$ should be no less than $\log_p(f/F)+1\sim14$ (it will yield around 10 random triggers per hour). For the target rate $f_t=1$~kHz $n$ will be about 11 (and about 2.5 random triggers per second). These values are relatively high. SPHERE-2 had 3--5 ph.e. thresholds that worked for big pixels. The small area of a SiPM pixel in the new detector means that fewer signal photons (from EAS) under comparable conditions will hit it, even with respect to the large aperture. This leads to an increase in the energy threshold of the detector. However, the new detector's target low energy threshold is, again, lower than that of SPHERE-2, thus the number of photons reaching each pixel from an event that should be registered is quite low, on the scale of 3--5.

To solve this problem several options exist. Each with their pros and cons.

Reduction of pixel activated state duration $\tau$ will reduce random coincidence rates but can altogether stop the trigger system from activating upon EAS events since they have a time structure measured on the order of microseconds. This includes the time that it takes the EAS plane to cross the observation plane plus differences in optical path lengths for different pixels. These values increase with altitude. 

\begin{figure}[tbp]
    \includegraphics[width=.45\textwidth]{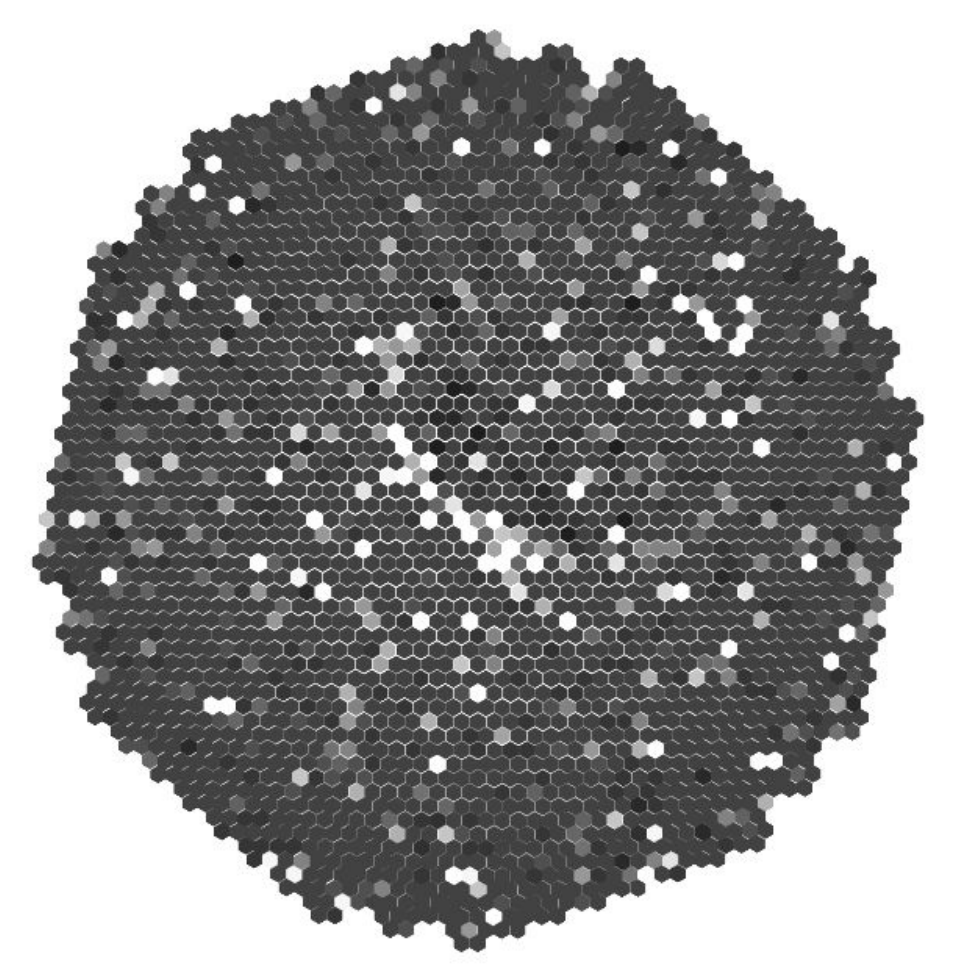}% 
    \caption{Example of an EAS CL structure on the mosaic from a primary 10~PeV iron nuclei. The image represents some instant signal values in pixels and not the total signal collected. The crescent shape of EAS is the result of photon arrival delays due to differences in optical path lengths.}
    \label{fig:img_03} 
\end{figure}

Lowering the thresholds by increasing $k$ --- the number of simultaneously activated pixels required for a trigger to work (random trigger rate is roughly $N_pf_t^k\tau^{k-1}$) --- will result in a topological complexity of defining which set of pixels counts as adjacent on a hexagonal grid. Also this will increase the number of missed events at lower energies --- a dense patch of pixels will only be activated by bright events, since fluctuations and distribution of the EAS CL photons has a complex structure on the mosaic. On Fig.~\ref{fig:img_03} an example of a signal from an EAS event is shown for reference. The image does not look even remotely smooth.

Implementation of online processing of complex visual patterns will require additional computing power and, critically more computational time than there reasonably is to make an `an event'/`not an event' trigger decision. This time is at most no more than 1--4 ticks of the data acquisition system clock or a bit more if some parallelization or data processing conveyor is designed, e.g. 4 parallel processing lanes will give 4 times more time to make a decision, but still this is a very limited time window.

\section{Method}
\label{sec:method}

To circumvent this limitation, it was proposed to use a two-stage trigger system. The idea is that the topological trigger itself is only the first stage of random noise filtration. After it another system checks that there is event-like data in the buffer. However, it should be noted that this second stage check should be done in a relatively fast manner. 

This filtration approach of random coincidence triggers allows to eased the overall trigger rate constrains, since there is a limitation only on the overall registration rate (from the data acquisition system system operation speed or limitation of available data storage space). In the first stage, a topological trigger is activated, and a small sample recording begins on a fixed number of bins. The data is then transferred to the second stage designed to filter out fragments without an EAS signal. Thus, the problem of binary classification is addressed.

Operation of the trigger's second stage requires the identification of complex visual patterns, for what convolutional neural networks (CNN)~\cite{CNN} were chosen. Section~\ref{subsec:data} describes the process of collecting and preparing data for training, and section~\ref{subsec:solution} describes the neural network and training parameters.

\subsection{Data Preparation}
\label{subsec:data}
To generate a dataset the process of EAS photons hitting the detector mosaic was simulated. This simulation consisted of 4 stages: generation of a bank of EAS events using the Monte Carlo simulation (with optical background estimation), modeling of the passage of reflected photons through the atmosphere, modeling of the passage of photons through the optical design of the telescope, electronics (SiPM) response calculation.

\begin{table}[tp]
    \begin{center}
        \caption{EAS modeling parameters.}
        \label{tab:corsika_params}
        \begin{tabular}{l r} 
            \hline
            Parameter name & Value \\ 
            \hline 
            hadron interaction model  & QGSJETII-04\\ 
            atmosphere model & No.1 (US. standard)\\
            %&&parameterized by Linsley \\
            telescope altitude above snow& 1000 m\\
            observation level & 450 m\\
            range of axis coordinates & $\pm500$~m \\
            \hline
        \end{tabular}
    \end{center}
\end{table}

\begin{table}[bp]
    \begin{center}
        \caption{SPHERE-3 telescope optical system parameters.}
        \label{tab:optics_params}
        \begin{tabular}{l c r} 
            \hline
            Parameter name & \hspace{8 mm} & Value \\ 
            \hline 
            curvature radius of the mirror && 1654 mm\\ 
            mosaic radius && 340 mm \\
            curvature radius of the mosaic && 868 mm\\
            aperture radius && 850 mm \\
            light collector lens radius && 7 mm \\
            entrance aperture area && $\sim2.27 \text{ m}^2$ \\ 
            \hline
        \end{tabular}
    \end{center}
\end{table}

\begin{table}[tp]
    \begin{center}
        \caption{Electronics response simulation settings.}
        \label{tab:electronics_params}
        \begin{tabular}{l c r} 
            \hline
            Parameter name  & \hspace{5 mm} & Value \\ 
            \hline            
            SiPM voltage && 29.6 V \\ 
            SiPM temperature && $-15.0 ^{\circ}$C \\ 
            SiPM overvoltage && 6.02 V\\
            %Gain distribution width && 0.82 \\ 
            Background photon amplitude && 0.013 ph/ns \\ 
            Digitization frequency && 80 MHz \\
            \hline
        \end{tabular}
    \end{center}
\end{table}

\begin{enumerate}
  \item For the Monte Carlo simulation, the CORSIKA~\cite{Heck1998} package was used. 100~events from 10~PeV primary iron nuclei with the same zenith angle (10 degrees) were simulated. For each simulated event, the coordinates of the shower axis relative to the telescope were also randomly selected 100~times, increasing the number of independent samples to 10\,000. Event parameters are given in table~\ref{tab:corsika_params}. 
  \item Simulation of the passage of photons from the snow to the detector through the atmosphere involved geometrical reprojecting of the light spot on the snow into a light spot on the entrance aperture of the detector.
  \item For simulation of the passage of photons through the optical design of the telescope, the Geant4~\cite{Geant4y2003} package was used. An optical design with a maximized entrance aperture was selected. The telescope geometry was built using STL files since the mirror and corrector plate was too complex for Geant4 in-built primitives. Configuration parameters are shown in the table~\ref{tab:optics_params}.
  \item Accounting for electronics involved modeling the response of the data acquisition system to the stream of photons. Since the system at this level is close to linear (operating far from the amplifiers' limits), the response was calculated as the sum of individual responses to each of the EAS' and background photons. The response accounts for SiPM cross-talks (parameters were taken from~\cite{Amineva2023} for the same SiPM type --- SensL MicroFC-SMTPA-60035~\cite{SensL}), amplification fluctuations, output pulse profile (see~\cite{YADFIZ2023}),  digitization process (including clock shift across different elements), and more. Electronics simulation parameters are given in table~\ref{tab:electronics_params}.
\end{enumerate}

\begin{figure}[tbp]
    \includegraphics[width=.45\textwidth]{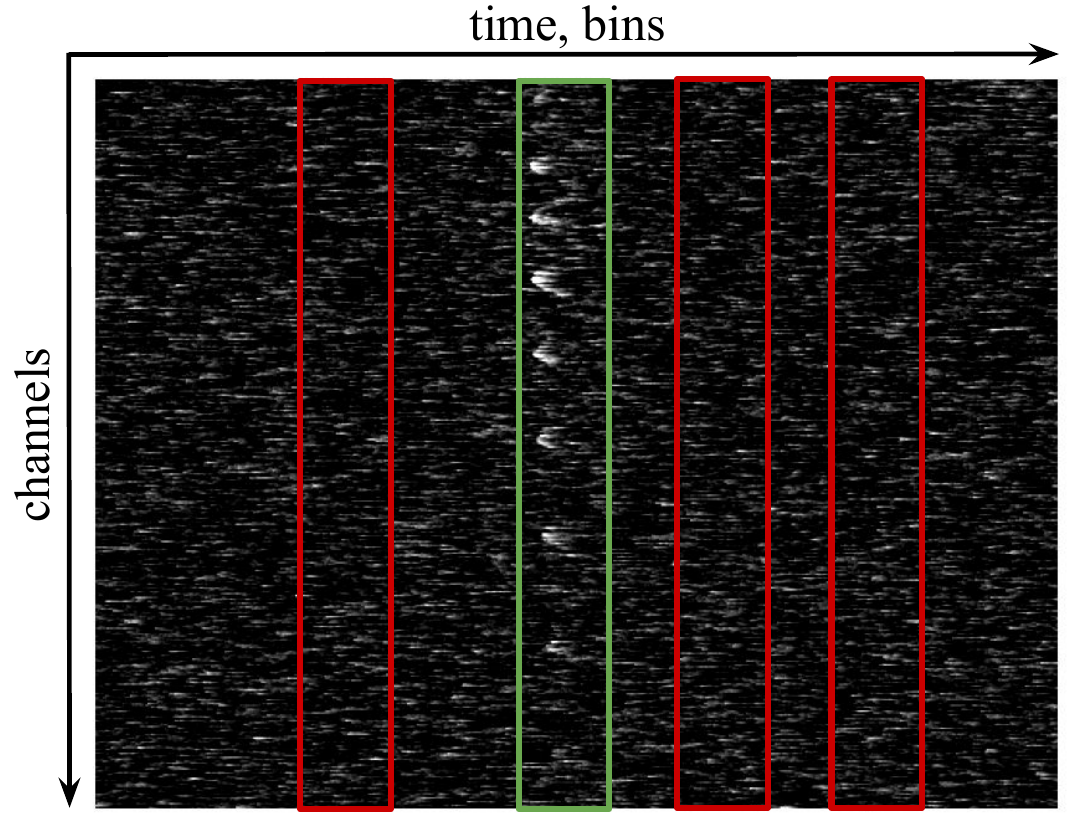}% 
    \caption{Dataset preparation step. The red highlighted areas contain only noise, and the green highlighted area contains the EAS-event signal.}
    \label{fig:img_04}
\end{figure}

The electronics output signal was a 500 bin long (6.25~$\mu$s) time sequence for each of the 2653 pixels, near the 225 time bin (almost the center) of which was the simulated event. For this 50 bin long non-overlapping fragments (625~ns) were cut out from the simulated sequence, some containing the full event (since the event location was known), some --- pure background, and each was labelled (Fig.~\ref{fig:img_04}). After applying all of the described modifications, the dataset was split into a train and test portions in a ratio of 8:2 and normalized. In total the training set contained 8\,000 samples and the test set contained 2\,000 samples, both with a 1:1 event to background ratio. Then the convolutional neural network was constructed and trained.

Working on pre-recorded sections also allows to use all of the information about a part of the time sequence. This allows to treat the time dimension as another spatial dimension, and thus use 2D convolutional layers.

\subsection{Implementation details}
\label{subsec:solution}

A practically minimal convolutional neural network architecture was chosen so as to fit if needed onto a microcontroller or FPGA chip. It consisted of 4 convolutional layers and one fully connected layer, as detailed in table~\ref{tab:nn_params}. The parameters column contains a line ($A$, $B$, $C\times{}D$, $E$), which should be interpreted as follows: $A$ is the number of input channels, $B$ - the number of output channels,  $C\times{}D$ is the kernel size, $E$ is the stride. The ReLU~\cite{ReLU} was used as a non-linear activation function, and the output layer was normalized using the softmax function, allowing the interpretation of logits as class probabilities. 

\begin{table}[h]
    \begin{center}
        \caption{Neural Network Architecture.}
        \label{tab:nn_params}
        \begin{tabular}{lc r} 
            \hline
            Layer Name  & \hspace{5 mm} & parameters \\ 
            \hline
            conv2d && (1, 2, 3$\times$3, 1) \\ 
            conv2d && (2, 6, 4$\times$4, 2) \\ 
            conv2d && (6, 3, 4$\times$4, 2) \\ 
            conv2d && (3, 3, 4$\times$4, 4) \\ 
            dense && (345 , 2), bias \\ 
            \hline
        \end{tabular}
    \end{center}
\end{table}

For training we used the negative log likelihood as a loss function. The Adam optimizer~\cite{Adam2014} was used with an initial learning rate of $10^{-3}$. The CNN was trained for a total of 50 epochs. For validation and evaluation of the final neural network, only accuracy  (number of correctly predicted labels) was used.

\section{Results}
\label{sec:results}

\begin{table}[bt]
    \begin{center}
        \caption{Classification results.}
        \label{tab:results}
        \begin{tabular}{|c|c|c|c|c|}
            \hline
            \multirow{3}{*}{real label}  & \multicolumn{4}{c|}{detected as} \\ 
            \cline{2-5} 
            & \multicolumn{2}{c|}{without threshold}& \multicolumn{2}{c|}{with threshold} \\ 
            \cline{2-5} 
            & True   & False & True   & False\\ 
            \hline
            True & 99.3\% & 0.7\% & 97.2\%  & 2.8\%  \\ 
            \hline
            False & 1.0\%  & 99.0\%  &0.1\% & 99.9\% \\ 
            \hline
        \end{tabular}
    \end{center}
\end{table} 

The classification accuracy is given in the table~\ref{tab:results}. The critical metric for the two-stage trigger system is the false-positive probability, since during the normal detector operation the event rate is $10^6$ times rarer than the expected noise rate. Without any additional manipulations with the outputs the neural network yields a 1.0\% false positive rate what is a good result. However, introduction of class separation thresholds allows to lower the false positive rate. The separation threshold is an additional constant that is applied to the neural network outputs and allows to artificially inflate or deflate the probability of predicting a specific class. Fine-tuning this value allows to achieve maximum filtering with an acceptable number of missed events. In table~\ref{tab:results} the ``without threshold'' columns show the metrics for the pure outputs of the neural network. The ``with threshold'' columns show the results with class separation threshold applying.

By adjusting the class separability threshold, the number of false positives can be reduced to 0.1\%, albeit at the cost of losing 2.8\% of events (false negative rate).

\section{Discussion}
\label{sec:discussion}

The comparison with SPHERE-2 was carried out in a somewhat complex way. The energy threshold of the SPHERE telescopes depends on the altitude (nearly linearly). Majority of the SPHERE-2 flights were carried out at around 480--500~m altitude. The estimated energy threshold for this altitude was around 10~PeV (see section~5.2~in~\cite{Antonov2015b} for details). The expected SPHERE-2 energy threshold for 1000~m altitude would've been around 20 PeV. 

\begin{figure}[tb]
    \includegraphics{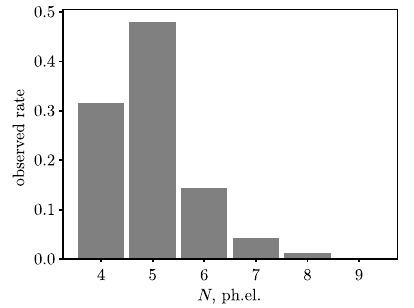}% 
    \caption{Distribution of the third pixel amplitude in the brightest triplet of the EAS image.}
    \label{fig:img_05}
\end{figure}

Simulations for SPHERE-3 in this work showed that 10~PeV EAS rarely produce amplitudes corresponding to 14 photoelectrons in the third brightest pixel of the whole event. However, in 50\% of the cases the ``bleakest'' pixel of the brightest triplet had around 5 photoelectrons (see Fig.~\ref{fig:img_05}). This means that direct application of the SPHERE-2 logic to the SPHERE-3 detector (as described in sec.~\ref{sec:problem}) would lead to an energy threshold of above 30~PeV (since EAS CL flux is proportional to primary energy). The 0.1\% false positive rate of the neural network obtained in the previous paragraph allows to have a 10$^3$ times higher rate of trigger activation. This pixel activation rate corresponds to a 6--7 photoelectrons amplitude (e.g. almost two times lower) and around 12--15~PeV detector energy threshold.

Precision of the primary particle parameters reconstruction in case of such low signal amplitudes is a matter of another study, however, the large number of pixels allows to use statistical approaches for data analysis and further detector operation optimization. Also, the plausibility of the neural network realization using limited computational power of FPGA chips or microcontrollers was not yet fully studied since no definite solutions were selected for detector electronics realization.

Other approaches to the trigger system logic (shorter pixel activation time, switching from triplets to septets and etc.) will be also studied for their effective energy thresholds. It also should be kept in mind, that this study was performed for reflected EAS CL that has a relatively long time structure. The option of direct CL registration by the main telescope camera was not included in the scope of this study, as it has different temporal properties. However, for its case the trigger system should also account for short bright bursts of direct CL as a factor of operation.

\section*{Acknowledgements}
The research was carried out using the equipment of the shared research facilities of HPC computing resources at Lomonosov Moscow State University~\cite{SC}.
This work is supported by the Russian Science Foundation under \href{https://rscf.ru/project/23-72-00006/}{Grant No. 23-72-00006}.

\bibliography{DLCP}

\end{document}